\documentclass[a4paper,12pt]{article}
\usepackage{amssymb}
\usepackage{amsmath}
\usepackage{epsfig}
\usepackage{xspace}
\usepackage{cite}

\newcommand{\eg}{e.g.\xspace}
\newcommand{\GeV}{\,\,\mathrm{GeV}}
\newcommand{\kbar}{{\bar k}}
\newcommand{\CG}{{\cal G}}
\newcommand{\CGO}{{\cal G}_{\omega}}
\newcommand{\CKO}{{\cal K}_{\omega}}
\newcommand{\ci}{{\mathrm{c}}}
\newcommand{\as}{\alpha_s}              
\newcommand{\asb}{\bar{\alpha}_s}       
\newcommand{\nf}{n_f}
\newcommand{\eff}{{\mathrm{eff}}}

\newcommand{\half}{{\textstyle \frac12}}
\newcommand{\kt}{{\boldsymbol k}}       
\newcommand{\om}{\omega}
\newcommand{\omhalf}{{\textstyle\frac{\omega}{2}}}
\newcommand{\omc}{\omega_c}
\newcommand{\omp}{\omega_\mathbb{P}}
\newcommand{\oms}{\omega_s}
\newcommand{\ord}[1]{\mathcal{O}\left(#1\right)} 

\newcommand{\NLLB}{$\mathrm{NLL_B}$\xspace}

\newcommand{\epjc}[3]{{\it Eur.~Phys.~J.~}{\bf C #1} (#2) #3}
\newcommand{\hep}[1]{{\tt hep-ph/#1}}
\newcommand{\jhep}[3]{{\it JHEP }{\bf #1} (#2) #3}
\newcommand{\jetp}[3]{{\it Sov.~Phys.~JETP }{\bf #1} (#2) #3}
\newcommand{\jetpl}[3]{{\it JETP Lett.~}{\bf #1} (#2) #3}

\newcommand{\npb}[3]{{\it Nucl.~Phys.~}{\bf B #1} (#2) #3}

\newcommand{\plb}[3]{{\it Phys.~Lett.~}{\bf B #1} (#2) #3}
\newcommand{\prd}[3]{{\it Phys.~Rev.~}{\bf D #1} (#2) #3}

\newcommand{\sjnp}[3]{{\it Sov.~J.~Nucl.~Phys.~}{\bf #1} (#2) #3}
\newcommand{\zpc}[3]{{\it Z.~Phys.~}{\bf C #1} (#2) #3}

\begin{document}     

\titlepage

\begin{flushright}
DESY 03--064 \\ DFF 403/05/03\\  LPTHE--03--15 \\ hep-ph/0305254 \\
\end{flushright}

\begin{center}
\vspace{2.5cm}
{\bf\Large Extending QCD perturbation theory to higher energies}
\vspace{0.7cm}

{\large \textsf{M.~Ciafaloni$^{(a)}$,
D.~Colferai$^{(a)}$,
G.~P.~Salam$^{(b)}$
and A.~M.~Sta\'sto$^{(c)}$}}\\
\vspace*{0.7cm}
$^{(a)}$ Dipartimento di Fisica, Universit\'a di Firenze,
50019 Sesto Fiorentino (FI), Italy;\\
INFN Sezione di Firenze, 50019 Sesto Fiorentino (FI), Italy\\
\vskip 2mm
$^{(b)}$ LPTHE, Universit\'es Paris VI and VII, CNRS UMR 7589, Paris
75005, France\\
\vskip 2mm
$^{(c)}$ Theory Division, DESY, D22603 Hamburg, Germany;\\
H.~Niewodnicza\'nski Institute of Nuclear Physics, Krak\'ow, Poland\\
\vskip 2cm
\end{center}
\thispagestyle{empty}

\begin{abstract}
  On the basis of the results of a new renormalisation group improved
  small-$x$ resummation scheme, we argue that the range of validity of
  perturbative calculations is considerably extended in rapidity with
  respect to leading log expectations.  We thus provide predictions
  for the energy dependence of the gluon Green function in its
  perturbative domain and for the resummed splitting function. As in
  previous analyses, high-energy exponents are reduced to
  phenomenologically acceptable values. Additionally, interesting preasymptotic
  effects are observed. In particular, the splitting function shows a
  shallow dip in the moderate small-$x$ region, followed by the expected
  power increase.
\end{abstract}

\vskip 2cm
\begin{center}
\today
\\[1cm]
\end{center}
\newpage


\section{Introduction\label{s:intro}}

The prediction of high-energy hard cross sections in QCD perturbation
theory has been a puzzling problem in recent years for a number of
reasons: the existence of large perturbative leading $\log s$ (LL)
contributions~\cite{BFKL} that seem incompatible with a range of
experimental data; the discovery of even larger subleading
contributions (NLL) of opposite sign~\cite{NLLFL,NLLCC};
and the increasing importance of low-$k_t$ partons at high energies
leading asymptotically to a strong-coupling Pomeron regime
which can at best be modelled but not
really predicted~\cite{JKCOL,Lipatov86,CC97,CC1}.

In order to tame the problem of large logarithms with alternating sign
several resummation procedures have been
proposed~\cite{Salam1998,CC,CCS1,ABF2000,ABF2001,THORNE,SCHMIDT,BFKLP}. The
renormalisation group improved (RGI) approach~\cite{Salam1998,CC,CCS1}
offers a general understanding of the large subleading $\log s$ terms
as due to 
leading $\log Q^2$ collinear contributions which can be extracted on
the basis of the perturbative anomalous dimensions. As a result,
resummed high-energy exponents have been calculated in a stable way.
In a forthcoming paper~\cite{CCSS3} this resummation approach is
extended to the gluon Green's function and splitting function so as to
provide a fairly complete account of their energy dependence and of
its perturbative (PT) versus non-perturbative (NP) aspects. The purpose of the
present paper is to summarise the main results of Ref.~\cite{CCSS3}
and to argue, on this basis, that the RGI approach tames -- to a large
extent -- the problem of the strong-coupling region as well.
This is because the resummation of the (alternating sign) large logarithms
leads to smaller high-energy exponents (and diffusion coefficients)
and to a considerable 
suppression of the non-perturbative Pomeron itself in the Green
function. Furthermore, we find that strong-coupling contributions
factorise, as expected, in the collinear limit, and we are able to
provide the resummed perturbative splitting function in $x$-space.

The basic problem that we consider is the calculation of the
(azimuthally averaged) gluon Green function $G(Y;k,k_0)$ as a function
of the magnitudes of the external gluon transverse momenta $k \equiv
|\kt|,\;k_0 \equiv |\kt_0| $ and of the rapidity $Y\equiv \log
\frac{s}{k k_0}$. This is not yet a hard cross section, because we
need to incorporate~\cite{CCSS3} the impact factors of the
probes~\cite{LOImpact,BaCoGiKy,Bartels02}.
Nevertheless, the Green function exhibits most of the physical
features of the hard
process, if we think of $k^2,\;k_0^2$ as external (hard) scales. The
limits $k^2\gg k_0^2$ ($k_0^2 \gg k^2$) correspond conventionally to
the ordered (anti-ordered) collinear limit. By definition, in the
$\om$-space conjugate to $Y$ (so that $\hat{\om} = \partial_Y$) we
set
\begin{align}
 \label{defGGF}
 \CGO(\kt,\kt_0) &\equiv [\om - \CKO]^{-1} (\kt,\kt_0) \,,  \\
 \label{eqGGF}
 \om \CGO(\kt,\kt_0) &= \delta^2(\kt-\kt_0) +
 \int d^2 \kt' \; \CKO(\kt,\kt') \CGO(\kt',\kt_0) \;,
\end{align}
where $\CKO(\kt,\kt')$ is a kernel to be defined, whose $\om = 0$
limit is related to the BFKL $Y$-evolution kernel, known at LL and NLL
levels~\cite{NLLFL,NLLCC}.

The RGI approach is based on the simple observation that, in BFKL
iteration, all possible orderings of transverse momenta are to be
included, the ordered (anti-ordered) sequence
$k \gg k_1 \cdots \gg k_n \cdots \gg k_0$
($k \ll k_1 \cdots \ll k_n \cdots \ll k_0$)
showing scaling violations with Bjorken variable $k^2/s$ ($k_0^2/s$).
Therefore, if only leading $\log k^2$ contributions were to be
considered, the kernel $\CKO$ acting on
$\frac{1}{k^2}\left(\frac{k^2}{k_0^2}\right)^\gamma$
would be approximately represented by the
following eigenvalue function (in the frozen coupling limit)
\begin{equation}\label{CKOform}
 \frac{1}{\om} \CKO \rightarrow \asb
 \left( \frac1{\gamma+\omhalf} + \frac1{1+\omhalf-\gamma} \right)
 \left( \frac1{\om} + A_1(\om) \right) + \cdots \;, \qquad
 \asb \equiv \as \frac{N_c}{\pi} \;,
\end{equation}
where $\gamma^{(1)}_{gg} = \asb \left( \frac1{\om} + A_1(\om) \right)$
is the one-loop gluon-gluon anomalous dimension and we have introduced
the variable $\gamma$ conjugate to $\log k^2$. Note, in fact, that
Eq.~(\ref{CKOform}) reduces to the normal DGLAP evolution~\cite{DGLAP}
in $\log k^2$ ($\log k_0^2$) in the two orderings mentioned before,
because $\gamma+\omhalf$ ($1+\omhalf-\gamma$) is represented by
$\partial_{\log k^2}$ ($\partial_{\log k_0^2}$) at fixed values of $x
= k^2 /s$ ($x_0 = k_0^2 /s$) in the ordered (anti-ordered) momentum
region. Note also the $\om$-dependent shift~\cite{Salam1998,CC,CCS1}
of the $\gamma$-singularities occurring in Eq.~(\ref{CKOform}), which
is required by the change of scale ($k_0^2$ versus $k^2$) needed to
interchange the orderings, i.e., $x_0$ versus $x$.

We thus understand that the $\om$-dependence of $\CKO$ is essential
for the resummation of the collinear terms and can be used to
incorporate the exact LL collinear behaviour, while on the other hand,
the $\om \to 0$ behaviour of $\CKO$ is fixed by the BFKL limit up to
$\ord{\om}$ terms, so as to incorporate exact LL and NLL kernels. Such
requirements fix the kernel up to contributions that are NNL in $\ln
x$ and NL in $\ln Q^2$.\footnote{We choose not to include the known
  exact NL $\ln Q^2$ terms because the conceptual issues, associated
  for example with factorisation scheme dependence, have yet to be
  fully understood.}  The form proposed in~\cite{CCSS3} (there called
\NLLB) is given by
\begin{equation}\label{kernel}
 \CKO = \left(\asb(x_\mu^2 q^2) + b\asb^2 \ln x_\mu^2 \right) K^\om_0
 + \om \left(\asb(x_\mu^2 k_>^2) + b\asb^2 \ln x_\mu^2 \right) K^\om_c
 + \asb^2(x_\mu^2 k_>^2) \tilde{K}^\om_1\,,
\end{equation}
where $q \equiv |\kt - \kt'|$, $k_> \equiv \max(k,k')$,
$k_< \equiv \min(k,k')$, $x_\mu$ has been introduced to test
renormalisation-scale dependence, and the scale-invariant kernels in
the RHS have the eigenvalue functions
\begin{subequations}\label{impKer}
\begin{align}
 \label{chi0}
  \chi_0^\om(\gamma) &=
  2 \psi(1) - \psi(\gamma+\omhalf) - \psi(1-\gamma+\omhalf)\,, \\
 \label{chicoll}
  \chi_{\ci}^\om(\gamma) &=
  \frac{A_1(\om)}{\gamma + \omhalf} + \frac{A_1(\om)}{1 - \gamma + \omhalf}\,, \\
 \nonumber
  \tilde{\chi}_1^{\om}(\gamma) &= \chi_1(\gamma)  
  + \frac12 \chi_0(\gamma) \frac{\pi^2}{\sin^2(\pi\gamma)} 
  -\chi_0(\gamma) \frac{A_1(0)}{\gamma(1-\gamma)}
  + \frac{b}{2} \left[\chi'_0(\gamma)+\chi_0^2(\gamma) \right] \\
 \label{chi1tilde}
  &\quad -\left( \frac1{\gamma} + \frac1{1-\gamma} \right) C(0) +
  \left( \frac1{\gamma+\omhalf} + \frac1{1+\omhalf-\gamma} \right)
  C(\om) [ 1+\om A_1(\om)]\,,
\end{align}
\end{subequations}
with $b = \frac{11}{12} - \frac{\nf}{6N_c}$,
$C(\om) = \frac{\psi(1+\om)-\psi(1)}{\om} - \frac{A_1(\om)}{\om+1}$ and
\begin{align}
 \label{chiLL}
  \chi_0(\gamma) &= \chi_0^{\om=0}(\gamma)
  = 2 \psi(1) - \psi(\gamma) - \psi(1-\gamma)\,, \\
 \nonumber
  \chi_1(\gamma) &= -\frac{b}{2} [\chi^2_0(\gamma) + \chi'_0(\gamma)] 
  -\frac{1}{4} \chi_0''(\gamma) 
  -\frac{1}{4} \left(\frac{\pi}{\sin \pi \gamma} \right)^2
  \frac{\cos \pi \gamma}{3 (1-2\gamma)}
  \left(11+\frac{\gamma (1-\gamma )}{(1+2\gamma)(3-2\gamma)}\right) \\
 \nonumber
  &\quad +\left(\frac{67}{36}-\frac{\pi^2}{12} \right) \chi_0(\gamma) 
  +\frac{3}{2} \zeta(3) + \frac{\pi^3}{4\sin \pi\gamma} \\
 \label{chiNLL}
  &\quad -\sum_{n=0}^{\infty} (-1)^n
  \left[ \frac{\psi(n+1+\gamma)-\psi(1)}{(n+\gamma)^2}
  +\frac{\psi(n+2-\gamma)-\psi(1)}{(n+1-\gamma)^2} \right] \;.
\end{align}
All $\nf$ dependence other than that in the running coupling terms
(that is in $b$) has been neglected. A correct treatment of all
$\nf$-dependent RG contributions would make the formalism technically
more complex (\eg requiring a two-channel approach) and given the
observed small effect of the $\nf$ contributions, we feel such an
effort to be currently uncalled for.

Note that, because of the explicit form of $\chi_0^\om$, the
kernel~(\ref{kernel}) reproduces the pole behaviour~(\ref{CKOform}) to
first order in $\as$. Note also that the last line in
Eq.~(\ref{chi1tilde}), which vanishes in the $\om = 0$ limit, has been
added in order to shift the remaining $C(0)/\gamma$ and
$C(0)/(1-\gamma)$ poles in $\chi_1$. There is some freedom in the
choice of the $\om$-dependence of the coefficient of the shifted
poles, and we take here a minimal prescription, one that gives an
indentically zero two-loop contribution to the anomalous dimension (we
refer to this prescription as \NLLB).  This also guarantees that the
momentum sum-rule is satisfied at two-loop level.  Note finally that
the resummed kernel proposed here differs from that of~\cite{CCS1}
because the collinear terms are added in the $\om$-dependent form, the
difference being at NNLL level. The reason for such a change is that
we have at most simple collinear poles, as in Eq.~(\ref{CKOform}), and
not double poles. This avoids the need for the $\om$ expansion, thus
providing a kernel more suitable for numerical iteration in
($x,k$)-space.

The integral equation to be solved by the definition~(\ref{eqGGF}) is
thus a running coupling equation with non linear dependence on $\as$
at appropriate scales, and it has a somewhat involved $\om$-dependence
in the improved kernels~(\ref{impKer}). Its solution has been found
in~\cite{CCSS3} by numerical matrix evolution methods in $k$- and $x$-
space~\cite{BoMaSaSc97}, where the typical $\om$-shifted form in the
example~(\ref{CKOform}) corresponds to the so-called consistency
constraint~\cite{Ciaf88,LDC,KMS1997}. Furthermore, introducing the
integrated gluon density
\begin{equation}\label{gluonDensity}
 x g(x,Q^2) \equiv \int^{Q^2} d^{2\!}\kt \; 
 G^{(s_0=k^2)}(\log 1/x; |\kt|, k_0) \;,
 \qquad \CG_\om^{(s_0=k^2)} \equiv \left(\frac{k}{k_0}\right)^\om \CGO \;,
\end{equation}
the resummed splitting function $P_\eff(z,Q^2)$ is defined by the
evolution equation
\begin{equation}\label{splitting}
 \frac{\partial g(x,Q^2)}{\partial \log Q^2} = \int \frac{dz}{z}\;
 P_\eff(z,Q^2)\; g\left( \frac{x}{z}, Q^2 \right)\,,
\end{equation}
and has been extracted~\cite{CCSS3} by a numerical deconvolution
method~\cite{CCS3}. We note immediately that $P_\eff$ turns out to be
independent of $k_0$ for $Q^2 \gg k_0^2$, yielding an important check
of RG factorisation in our approach.

\section{Gluon Green's function\label{s:ggf}}

\begin{figure}[tbp]
  \centering
  \includegraphics[width=0.4825\textwidth]{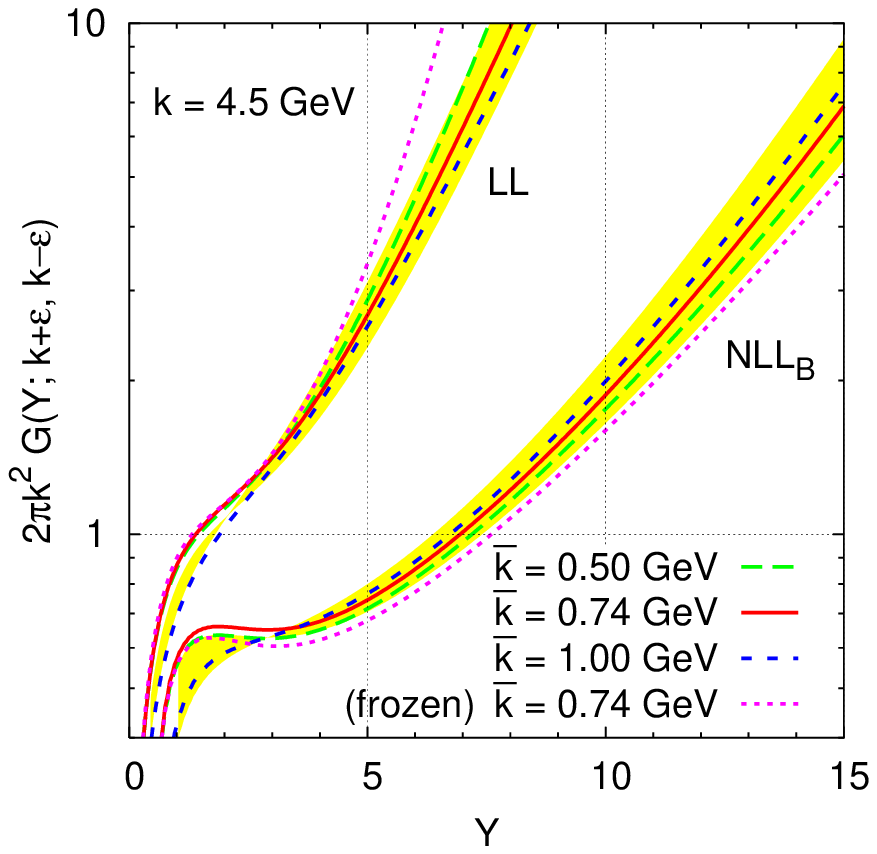}\hfill
  \includegraphics[width=0.50\textwidth]{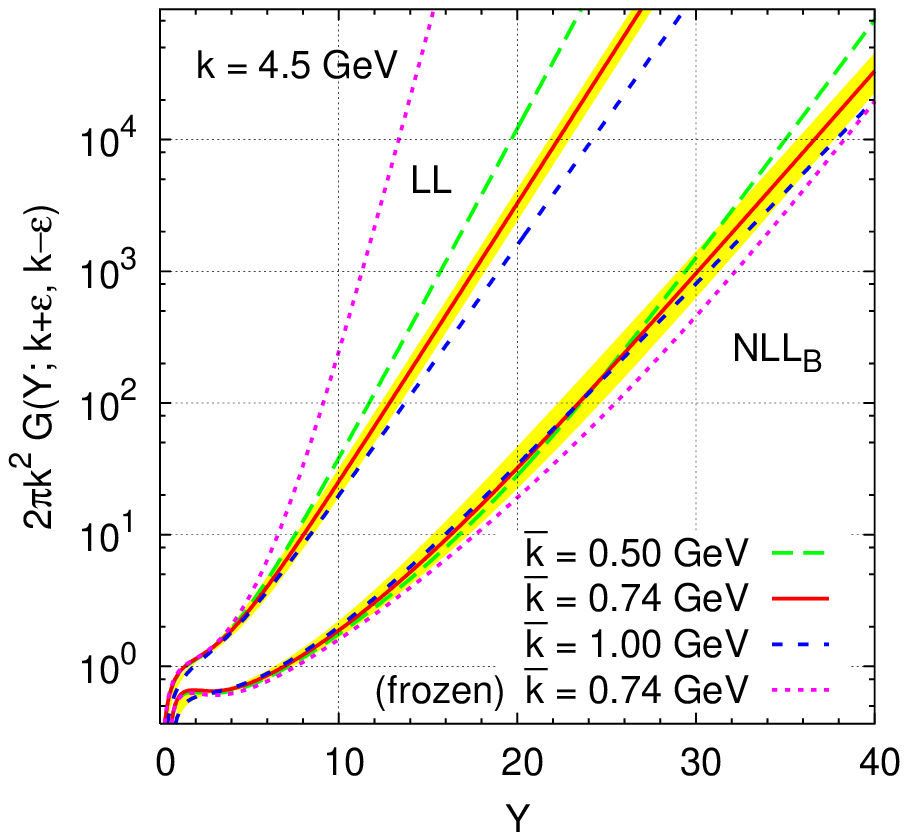}
  \caption{\it Green's function calculated with four different
    infrared regularisations of the coupling, shown for LL and RGI NLL
    (`NLL$_\mathrm{B}\!$') evolution. The bands indicate the
    sensitivity of the $\kbar=0.74\GeV$ results to a variation of
    $x^2_\mu$ in the range $\frac12$ to $2$. The left and right hand
    plots differ only in their scales.
  }
  \label{f:manyReg}
\end{figure}

Results for $G(Y;k,k)$%
\footnote{Actually, we take slightly different values of the scales in
  $G$, namely we consider $G(Y;k+\epsilon,k-\epsilon)$ with $\epsilon
  = 0.1 k$, in order to avoid sensitivity to the discretisation of the
  $\delta$-function initial condition in \eqref{eqGGF} (cf.\ 
  Ref.~\cite{CCSS3} for a detailed discussion).}  are shown in
Fig.~\ref{f:manyReg}.
In addition to the solution based on our RGI approach (\NLLB) the
figure also has `reference' results for LL evolution with kernel
$\asb(x_\mu^2 q^2) K^0_0$. We use a one-loop running coupling with
$\nf=4$, which is regularised either by setting it to zero below a
scale $\kbar$ (`cutoff') or by freezing it below that scale
($\as(q^2<\kbar^2) = \as(\kbar^2)$). We believe the cutoff
regularisation to be more physical since it prevents diffusion to
arbitrarily small scales and is thus more consistent with confinement ---
accordingly we show three cutoff 
regularisations and only one frozen regularisation. The $\kbar =
0.74\GeV$ cutoff solution is presented together with an uncertainty
band associated with the variation of $x^2_\mu$ between $\frac12$ and
$2$.

Solutions of \eqref{eqGGF} with an IR-regularised coupling generally
have two domains~\cite{JKCOL,Lipatov86,CC97,CCSS2},
separated by a critical rapidity $Y_c(k^2)$. For the
intermediate high-energy region $1 \ll Y \lesssim Y_c(k^2)$, one
expects the perturbative `hard Pomeron' behaviour with exponent $\oms$,
\begin{equation}\label{Gpert}
 k^2 G(Y;k,k) \sim \frac1{\sqrt{Y}}
 \exp \big[ \oms(\as(k^2)) Y + \Delta(\as, Y) \big] \;,
\end{equation}
and diffusion corrections~\cite{KOVMUELLER,ABB,LEVIN,CMT}
parametrised by $\Delta(\as, Y)$. Beyond $Y_c$, a
regularisation-dependent non-perturbative `Pomeron' regime takes over
\begin{equation}\label{Gnonpert}
 k^2 G(Y;k,k) \sim \left(\frac{\kbar^2}{k^2}\right)^{\xi} e^{\omp
   Y}\,,\qquad 
   \begin{array}{rl}
     \mathrm{LL}\;\,\,: & \xi = 1 \\ 
     \mathrm{NLL}_B: & \xi = 1 + \omp 
   \end{array}
%
\end{equation}
where the factor $\xi$ differs from $1$ only for kernels involving the
consistency constraint. The non-perturbative exponent $\omp$
satisfies~\cite{CCS1} $\omp\sim 
\oms(\as(\kbar^2))$ and hence
is formally larger than $\oms(\as(k^2))$.\footnote{The behaviour
  \eqref{Gnonpert} with $\omp > \oms(\as(k^2))$ is a general feature
  of linear evolution equations such as \eqref{eqGGF}, but not of actual
  high energy cross sections, which are additionally subject to
  non-linear effects and confinement.} %

The value of $Y_c$ depends strongly on $k$. In the tunnelling
approximation, it can be roughly estimated by equating
eqs.~(\ref{Gpert}) and (\ref{Gnonpert}) to yield~\cite{CCS2,CCSS2},
for any given regularisation procedure,
\begin{equation}\label{Yc}
 Y_c(k^2) \simeq \frac{\xi\log(k^2/\kbar^2)}{\omp - \om_s(\as(k^2))} \;,
\end{equation}
(again with $1+\omp \to 1$ for LL), showing an approximately linear increase
of $Y_c$ with $\log k^2$.

Within this logic, several aspects of Fig.~\ref{f:manyReg} are worth
commenting. The most striking feature of the LL evolution is its
strong dependence on the non-perturbative regularisation, even for
rapidities as low as $5$. The exact value of $Y_c$ depends on the
regularisation being used, ranging between $5$ and $10$.
In contrast, \NLLB evolution remains under perturbative control up to
much larger rapidities and the NP pomeron behaviour takes over only
for $Y \gtrsim 25$, where the three cutoff solutions start to
diverge. Regularisation dependence is also present at lower
rapidities. This seems to be due to power corrections to
\eqref{Gpert}, associated with the use of a coupling $\as(q^2)$
\cite{CCSS3}. The non-perturbative regime of the IR frozen coupling solution
is reached only later ($Y\gtrsim 30$), at the point where it starts to
grow more rapidly than the cutoff solutions.
A final point to note is renormalisation scale uncertainties of our
resummed results are sizeable -- of the order of several tens of
percent for $Y>4$ -- but seem anyhow quite modest compared to the
order(s) of magnitude difference with LL.\footnote{The $x_\mu$
  dependence of the LL solution is somewhat smaller than for \NLLB ---
  this may seem surprising, but at larger $Y$ the LL solution is in
  the NP domain, where non-linearities (in $\as$) reduce the $x_\mu$
  dependence.}

The fact that $Y_c$ is considerably larger for \NLLB
evolution is natural --- it is an expected consequence (see
Eq.\eqref{Yc}) of the fact that subleading corrections lower both the
PT and NP exponents. What is quite non-trivial however is that at
their respective $Y_c$'s the \NLLB Green's function is an order of
magnitude larger than the LL one: the subleading corrections increase
the overall amount of BFKL growth remaining within perturbative
control. This is due to a number of factors, among them a strong
reduction of the diffusion coefficient (see below).  We note that the
increase in the amount of perturbatively calculable growth is of particular
interest for the theoretical question (see \eg \cite{KOVMUELLER}) of
whether it is possible to perturbatively generate a high-density
gluonic system.

\begin{figure}[t]
  \newcommand{\figfrac}{0.47} \centering
  \includegraphics[height=\figfrac\textwidth]{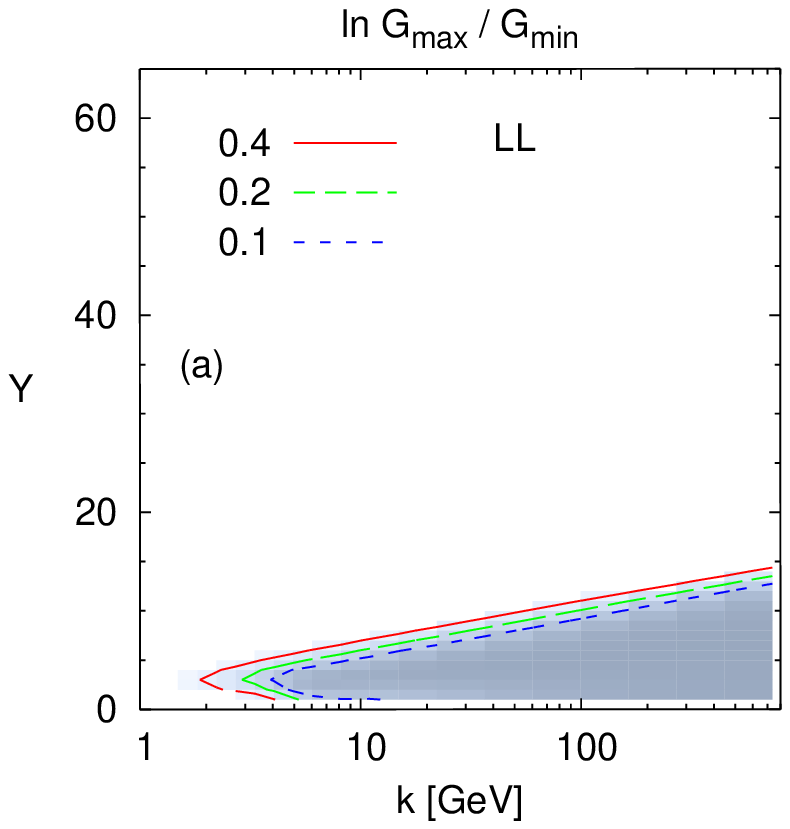} \hfill
  \includegraphics[height=\figfrac\textwidth]{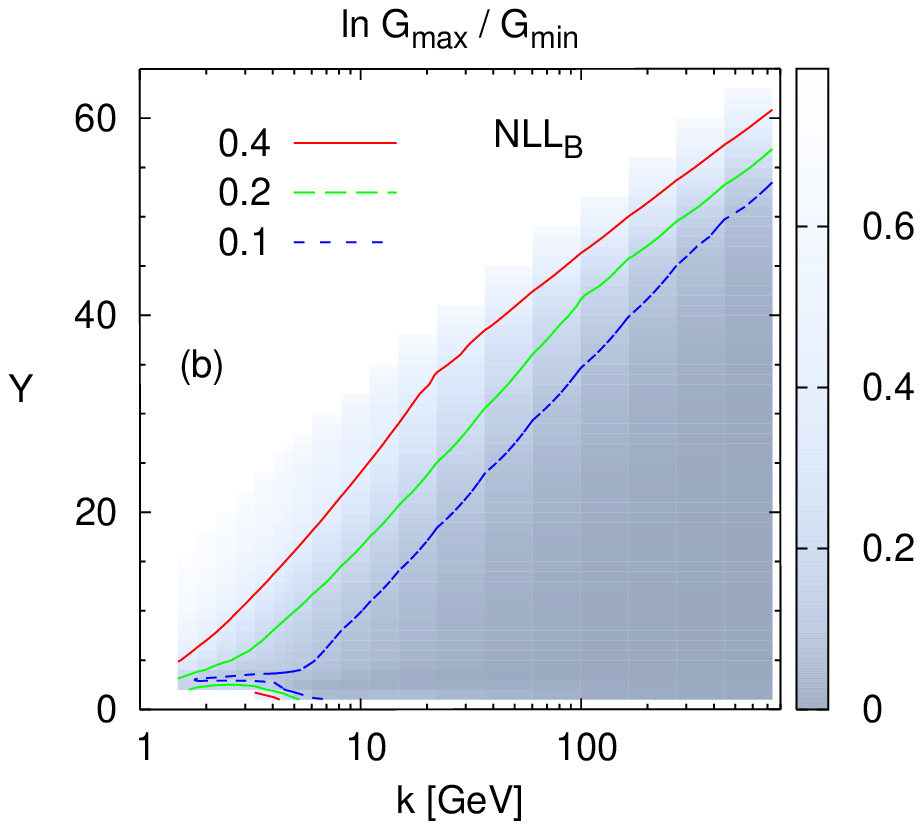}
  \caption{\it Contour plots showing the sensitivity of
    $G(Y,k+\epsilon, k-\epsilon)$ to one's choice of non-perturbative
    regularisation, as obtained by examining the logarithm of the ratio
    of the regularisations giving the largest and smallest result for
    $G$. Darker shades indicate insensitivity to the NP
    regularisation, and contours have been drawn where the logarithm
    of the ratio is equal to $0.1$, $0.2$ and $0.4$. Plot (a) shows
    the result for LL evolution, while (b) shows RGI NLL evolution
    (\NLLB). The regularisations considered are those of
    Fig.~\ref{f:manyReg}.}
 \label{f:valRegions}
\end{figure}

In Fig.~\ref{f:manyReg} we have considered only a single value of $k$.
The question of NP contributions is summarised more generally in
Fig.~\ref{f:valRegions}, which shows contour plots of the logarithmic
spread of our four regularisations. Darker regions are less IR
sensitive, and contours for particular values of the spread have been
added to guide the eye. Here too one clearly sees the much larger
region (including most of the phenomenologically interesting domain)
that is accessible perturbatively after accounting for subleading
corrections.

So let us now therefore return to Fig.~\ref{f:manyReg} and examine the
characteristics of the \NLLB Green's function in the perturbatively
accessible domain, which should be describable by an equation of the
form \eqref{Gpert}. The first feature to note is that the growth
starts only from $Y\simeq 4$. This suggests that at today's
collider energies (implying $Y\lesssim 6$~\cite{OPAL,L3}), it will at
best be possible to see only the start of any growth. This
preasymptotic feature is partly due to the slow opening of small-$x$
phase space~\cite{NONASYMP} implicit in our $\om$--shifting procedure.

Once the growth sets in,
the issue is to
establish the value of $\oms$ appearing in \eqref{Gpert}.  This is a
conceptually complex question because in contrast to the
fixed-coupling case, $\oms$ no longer corresponds to a Regge
singularity. There are running-coupling diffusion corrections
$\Delta(\as, Y)$~\cite{KOVMUELLER,ABB,LEVIN,CMT}, whose leading
contribution, $\sim Y^3$,  for our model is~\cite{CCSS3}
\begin{equation}\label{diffCorr}
 \Delta(\as, Y) \simeq  \frac{Y^3}{24}
 \left[ \frac{\partial}{\partial \log k^2}\oms(\as(k^2)) \right]^2
 \chi_\eff ''(\half)\,.
\end{equation}
In addition, $\Delta(\as, Y)$ contains terms with weaker $Y$
dependences, including $Y^2$ and $Y$.  Such terms can be disentangled
by the method of the $b$-expansion~\cite{CCSS1}.\footnote{ In the
  $b\to0$ limit, with $\as(k^2)$ kept almost fixed, the
  non-perturbative Pomeron is exponentially
  suppressed~\cite{CCSS1,ABF2001}, so that the
  $b$-expansion can also be used as a way of defining a purely
  perturbative Green's function without recourse to any
  particular infrared regularisation of the coupling~\cite{CCSS1}.} %
\begin{figure}[t]
\centering
\includegraphics[height=0.5\textwidth]{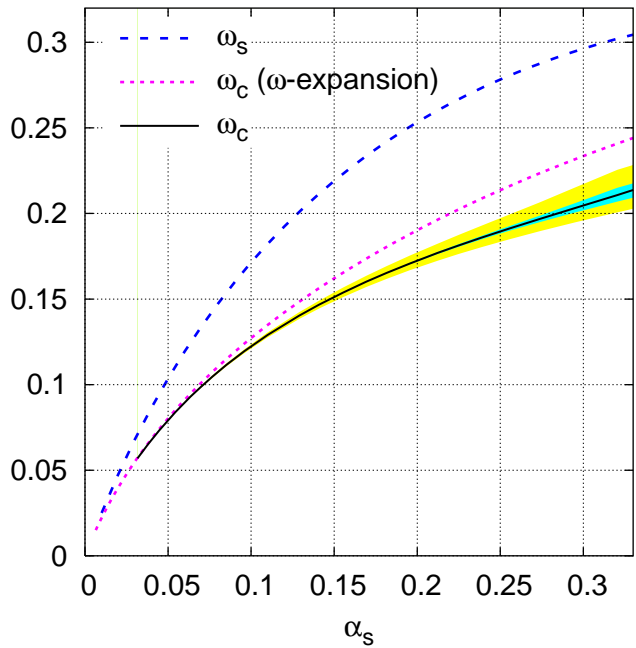}\;
\caption{\it The small-$x$ exponents: the Green's function effective exponent
  $\oms$ is shown to first order in the $b$-expansion; the splitting
  function exponent $\omc$ is shown together with NP and
  renormalisation scale uncertainty bands, defined in
  figure~\ref{f:Peff}. Also shown, for reference, is the result for
  $\omc$ using the method of~\cite{CCS1}, for $b(\nf = 4)$.}
\label{f:omc}
\end{figure}%
Since running-coupling diffusion corrections start only at order
$b^2$, it is possible to give an unambiguous definition of $\oms$ up
to first order in $b$, while retaining all orders in $\as$ for non
running-coupling effects. The result is shown in Fig.~\ref{f:omc} as a
function of $\as$ and as has been found in previous work~\cite{CCS1},
there is a sizeable decrease with respect to LL
expectations.\footnote{We do not compare directly with our earlier
  results for $\oms$~\cite{CCS1}, because they are based on a
  different definition (the saddle-point of an effective
  characteristic function), which is less directly related to the
  Green's function. Nevertheless, the present results are consistent with
  previous ones to within NNLL uncertainties.} %
Furthermore, the leading diffusion corrections in \eqref{diffCorr}
turn out to be numerically small, about an order of magnitude down
with respect to the LL result, due to a sizeable decrease of the
diffusion coefficient $\chi_\eff''$, over and above the decrease
already discussed for $\oms$.

A final point related to the Green's function concerns predictions
based on a pure NLL kernel without renormalisation group improvement.
Despite the fact that the characteristic function around
$\gamma=\frac12$ is very poorly convergent, it has been
argued~\cite{ROSS98} that the Green's function may nevertheless show a
growth corresponding to an effective positive $\oms$, due to
saddle-points at complex $\gamma$.  Indeed, we find~\cite{CCSS3} that
for $k \sim k_0$ the pure NLL Green's function is remarkably similar
to the \NLLB result and it is stable with respect to $x_\mu$
variations. There is a difference in normalisation, which we ascribe
to to the (implicit) presence of an effective impact factor in the
\NLLB solution~\cite{CCSS3}. It should be kept in mind however that
this `good behaviour' of the pure NLL Green's function breaks down
when $k$ and $k_0$ are substantially different (for $\asb \simeq 0.2$
the Green's function becomes negative for $|\ln k/k_0| \gtrsim 2$),
resulting in the unphysical oscillating behaviour predicted
in~\cite{ROSS98}.

\section{Resummed gluon splitting function\label{s:rgsf}}

Green's functions $G(Y;k,k_0)$ with $k_0 \sim \kbar \sim
\Lambda_\mathrm{QCD}$ are very sensitive to, and largely determined
by non-perturbative physics associated, in our numerical solutions,
with an IR regularisation of the running coupling at some scale
$\kbar$. For example at $x \equiv e^{-Y} = 10^{-10}$ and $k=4.5\GeV$,
the three cutoff regularisations of the previous section lead to a
spread of a factor of $5$ in calculations of the Green's function, in
sharp contrast to the good perturbative control seen in the previous
section, for the case $k_0\sim k \gg \Lambda_\mathrm{QCD}$.

\begin{figure}[ht]
\centering\includegraphics[width=0.70\textwidth]{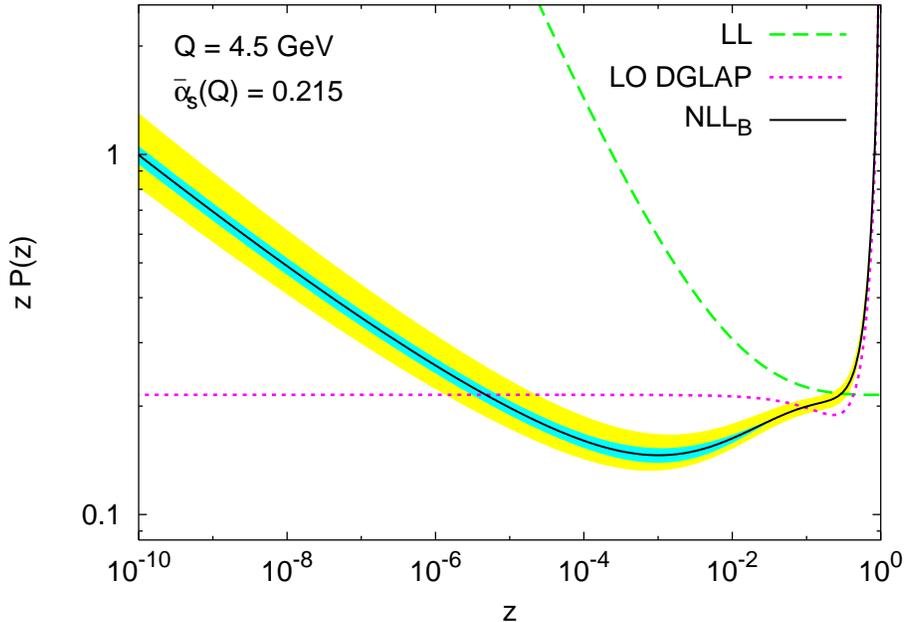}
\caption{\it Small-$x$ $\mathrm{NLL_B}$ resummed splitting function,
  compared to the pure 1-loop DGLAP and the (fixed-coupling) LL BFKL
  splitting functions. The central $\mathrm{NLL_B}$ result corresponds
  to $x_\mu=1$, $\kbar=0.74\GeV$; the inner band is that obtained by
  varying $\kbar$ between $0.5$ and $1.0\GeV$, while the outer band
  corresponds to $\frac12 < x_\mu^2 < 2$.}
\label{f:Peff}
\end{figure}

However, many arguments in the BFKL
framework,\cite{Lipatov86,JKCOL,CC1,CCS1,ABF2001,CCS3}, have been
given in favour of \emph{factorisation}, Eq.~\eqref{splitting}, with
the small-$x$ splitting function $P_{gg}(z,Q^2)$ being independent of
the IR regularisation.  The most dramatic demonstration of
factorisation is perhaps in the fact that a numerical extraction of
the splitting function from the Green's function by deconvolution
gives almost identical splitting functions regardless of the
regularisation. This is illustrated in Fig.~\ref{f:Peff}, 
where the
solid line and its inner band represent the result of the
deconvolution together with the uncertainty resulting from the
differences between the three cutoff regularisations.\footnote{Because
  of numerical instabilities in the rather delicate deconvolution
  procedure, we have so far not succeeded in obtaining reliable
  results with an IR frozen coupling --- preliminary results suggest
  however that the difference between frozen and cutoff
  regularisations is of the same order as the width of the inner band.} %
The resulting regularisation dependence is pretty small, and at higher
$Q$ it diminishes rapidly as an inverse power of $Q$, as expected from
a higher twist effect.

Several features of the resummed splitting function are worth
commenting and comparing with previous NLL calculations, including
various types of resummations~\cite{CCS1,ABF2001,THORNE}. Firstly, at
very large-$x$ it approaches the normal DGLAP splitting function. The
momentum sum rule is satisfied to within a few parts in $10^4$. At
moderately small-$x$ the splitting function is quite strongly
suppressed with respect to the LL result and shows not a power growth,
but instead a significant dip (of about $30\%$ relative to the LO
DGLAP value, for $\asb = 0.215$). Dips of
various sizes and positions have been observed before
in~\cite{THORNE,ABF2000,CCS3} though ours is significantly shallower
than that found in~\cite{THORNE} at NLL order and similar to that
found in the $\om$-expansion~\cite{CCSS3}.  An interesting question
concerns the impact of the dip on fits to parton distributions.
Calculations in a (partially) RGI LL model~\cite{KMS1997} whose
effective splitting function also has a dip, suggest that it is not
incompatible with the available structure function data.

At very small-$x$ one finally sees the BFKL growth of the splitting
function. We recall that the branch cut, present for a fixed
coupling, gets broken up into a string of poles, with the rightmost
pole located at $\omc$, to the left of the original branch point
($\oms$), $\oms - \omc \sim b^{2/3} \as^{5/3}$~\cite{CCS1}. The origin
of this correction is similar to that of the $b^{2/3} \as^{5/3}$
contributions to $\omp$ for cutoff regularisations~\cite{HR92}. The
dependence of $\omc$ on $Q$ is shown in Fig.~\ref{f:omc} together with
its scale and IR regularisation dependence. It is slightly lower than
the earlier determination in the
$\om$-expansion~\cite{CCS1}\footnote{when compared with the same
  flavour treatment ---
  the value of $\omc$ in Fig.~6 of~\cite{CCS1} actually refers to $b(\nf = 0)$,
  while that of Fig.~\ref{f:omc} here is for $b(\nf = 4)$.}.  Both
determinations are substantially below $\oms$, as expected.

\section{Conclusions}

In this letter we have outlined an approach to renormalisation-group
improved NLL BFKL resummation that is convenient for numerical
determination of the high-energy Green's functions and splitting
functions. This is an important step on the way to complete RGI NLL
predictions for high-energy cross sections and small-$x$ structure
function evolution.

We have discussed several important new results obtained within this
approach. Most striking is the increase in the domain in $k$ and $Y$
that becomes perturbatively accessible once one includes subleading
corrections. Even the amount of growth (number of `e-folds') of the
Green's function that can be calculated perturbatively increases
significantly.

Another result concerns the size of preasymptotic effects: for
example, for the transverse scale studied in Fig.~\ref{f:manyReg},
BFKL growth sets in only for rapidities greater than $4$. Hence it is
vital to study the full Green's function rather than just the
high-energy exponents discussed so far in the literature, and to include the
physical impact factors~\cite{BaCoGiKy,Bartels02} along the lines suggested
in\cite{CCSS3}.

In the collinear region, a key result is the practical demonstration of
factorisation of the small-$x$ Green function, and the extraction of
the resummed $P_{gg}$ splitting function. Here we obtain the
high-energy exponent $\omc$, but we find that preasymptotic effects
are again of fundamental phenomenological importance. As has been
observed in an approach without renormalisation-group
improvement~\cite{THORNE}, the main feature in today's accessible $x$
range is a small-$x$ \emph{dip} rather than growth (our dip is however
much shallower). This phenomenon has yet to be fully understood,
because several subleading effects are likely to come into play.

On the whole, the present work encourages us to trust resummed
perturbative calculations for next generation accelerators, and shows
that subleading contributions not only decrease high-energy
exponents, but provide significant preasymptotic effects.

\section*{Acknowledgements}
This work was supported in part by the Polish Committee for Scientific
Research (KBN) grants no.\ 2P03B~05119, 5P03B~14420 and by MURST
(Italy).  We also wish to thank the referee for the careful reading of
the text and useful suggestions.


\end{document}